\documentclass[aps,showpacs,preprintnumbers,amsmath, amssymb]{revtex4}

\usepackage{float}
\usepackage{graphics,epsfig}
\usepackage{graphicx}
\usepackage{dcolumn}
\usepackage{bm}

\begin{document}
\baselineskip=0.8 cm

\title{{\bf No-go theorem for static boson stars with negative cosmological constants }}
\author{Yan Peng$^{1}$\footnote{yanpengphy@163.com}}
\affiliation{\\$^{1}$ School of Mathematical Sciences, Qufu Normal University, Qufu, Shandong 273165, China}

\vspace*{0.2cm}
\begin{abstract}
\baselineskip=0.6 cm
\begin{center}
{\bf Abstract}
\end{center}

In a recent paper, Hod has proven no-go theorem for
asymptotically flat static regular boson stars.
In the present work, we extend discussions to the gravity with a negative
cosmological constant. We consider a scalar field vanishing at infinity.
In the asymptotically AdS background, we show that spherically symmetric
regular boson stars cannot be constructed
with self-gravitating static scalar fields, whose
potential is positive semidefinite and
increases with respect to its argument.

\end{abstract}

\maketitle
\newpage
\vspace*{0.2cm}

\section{Introduction}

There is accumulating evidence that fundamental scalar fields may exist in nature.
Black hole no hair theorems, see e.g. \cite{Bekenstein,Chase,C. Teitelboim,Ruffini-1},
play an important role in the development of the black hole theory.
Classical black hole no hair theorems
state that asymptotically flat black holes
cannot support static scalar fields outside the horizon,
see recent progress \cite{Hod-2}-\cite{Brihaye} and reviews \cite{Bekenstein-1,CAR}.

In contrast, it has recently been shown that
rotating black holes allow
the existence of stationary massive scalar field hairs \cite{CH0}-\cite{YW}.
Similarly, no static scalar hair behaviors were also found in horizonless reflecting
object backgrounds and rotating regular reflecting
objects can support stationary scalar hairs \cite{Hod-6}-\cite{LR1}.
A well known regular scalar configuration is the
boson star, which may theoretically be
described by either static or stationary scalar fields.
It was found that stationary self-gravitating massive scalar fields can
form the spatially regular boson star \cite{FE,DAER}.

Then whether static scalar fields can make boson stars
is a question to be answered. In the flat spacetime,
boson stars cannot be constructed with
static scalar fields due to Derrick¡¯s theorem \cite{GHD}.
Lately, Hod extended this no-go theorem for boson stars to
the asymptotically flat curved spacetime, considering
self-interaction static scalar fields possessing
a positive semidefinite potential increasing as a function of its argument \cite{ng1}.
It was further shown that this intriguing no-go theorem also holds for static
scalar fields nonminimally coupled to the asymptotically flat gravity \cite{ng2}.
On the other side, the AdS boundary could provide the confinement of the scalar field
and usually makes the scalar field easier to condense \cite{ng3,ng4,ng5}.
So it is of some interest to extend the discussion of no-go theorem
for static boson stars to the asymptotically AdS gravity.

In the following, we introduce the model
of self-gravitating static scalar fields
in the background with negative cosmological constants.
We prove that static scalar fields cannot form
spatially regular spherically symmetric boson stars
in the asymptotically AdS gravity.
We give conclusions in the
last section.

\section{No-go theorem for asymptotically AdS static boson stars}

We study the gravity system of static scalar field
in the spacetime with negative cosmological constants.
The Lagrangian density describing static scalar fields
in the curved spacetime reads \cite{dg,sh,Rogatko1}
\begin{eqnarray}\label{lagrange-1}
\mathcal{L}=\frac{R-2\Lambda}{16\pi G}-|\nabla_{\alpha} \psi|^{2}-V(\psi^{2}).
\end{eqnarray}
R is the Ricci curvature and $\Lambda<0$ is the negative cosmological constant.
Hereafter we choose $G=1$ for simplicity.
We take static scalar fields only depending on the
radial coordinate in the form $\psi=\psi(r)$.
The scalar field self-interaction potential $V(\psi^2)$ satisfies relations
\begin{eqnarray}\label{lagrange-1}
V(0)=0~~~~and~~~~\dot{V}=\frac{dV(\psi^2)}{d(\psi^2)}> 0.
\end{eqnarray}
It means the potential is positive semidefinite
and increases as a function of its argument.
And in the case of free scalar fields with mass $\mu$,
there is $V(\psi^2)=\mu^2\psi^2$, $V(0)=0$ and $\dot{V}=\mu^2>0$.

The four dimensional spherically symmetric boson star metric reads \cite{mr1,mr2,fc,Basu,Rogatko,Peng Wang}
\begin{eqnarray}\label{AdSBH}
ds^{2}&=&-fe^{-\chi}dt^{2}+\frac{dr^{2}}{f}+r^{2}(d\theta^2+sin^{2}\theta d\phi^{2}).
\end{eqnarray}
$\chi$ and $f=1-\frac{2m(r,\Lambda)}{r}$ are functions
depending on the radial coordinate r and cosmological constants,
where $m(r,\Lambda)$ is the effective mass \cite{SA,SA1}.
Angular coordinates are taken to be $\theta\in [0,\pi]$
and $\phi\in [0,2\pi]$.

We take the cosmological constant term $-\frac{\Lambda}{8\pi G}$
in the lagrange density as effective matter fields.
So both scalar fields and cosmological constants
contribute to the total energy density $\rho=-T_{t}^{t}$.
There is the relation
\begin{eqnarray}\label{InfBH}
\rho=\rho_{1}+\rho_{2},
\end{eqnarray}
where $\rho_{1}$ is the scalar field energy density
and $\rho_{2}$ is the energy density of cosmological constants.
The Einstein equations describing motions of
the matter field and the background is $G^{\mu}_{\nu}=8\pi T^{\mu}_{\nu}$.
It yields metric equations
\begin{eqnarray}\label{InfBH}
f'=-8\pi r \rho +(1-f)/r,
\end{eqnarray}
\begin{eqnarray}\label{InfBH}
\chi'=-8\pi r (\rho+p)/f,
\end{eqnarray}
where $p=T_{r}^{r}$ is the radial pressure \cite{mr2,fc,SA,SA1}.

Putting $f=1-\frac{2m(r,\Lambda)}{r}$ into (5),
we obtain the equation
\begin{eqnarray}\label{InfBH}
\frac{d m(r,\Lambda)}{dr}=4\pi r^{2}\rho,
\end{eqnarray}
which implies that \cite{SA,SA1}
\begin{eqnarray}\label{InfBH}
m(r,\Lambda)=\int_{0}^{r}4\pi r'^{2}\rho dr'.
\end{eqnarray}

The scalar field energy density reads
\begin{eqnarray}\label{InfBH}
\rho_{1}(r,\Lambda)=f(\psi')^2+V(\psi^2)
\end{eqnarray}
and the scalar field mass within a sphere with the radius r is given by
\begin{eqnarray}\label{InfBH}
m_{1}(r,\Lambda)=\int_{0}^{r}4\pi r'^{2}\rho_{1}(r',\Lambda)dr'.
\end{eqnarray}
Since (10) is a volume integral in the curved spacetime,
the integral should depends on the geometry (3).
In fact, $\rho_{1}$ depends on the metric
function f according to relation (9).
Within a sphere, $m_{1}(r,\Lambda)$ is the mass
corresponds to the energy density $\rho_{1}$,
which is due to matter field terms $-|\nabla_{\alpha} \psi|^{2}-V(\psi^{2})$
in the Lagrangian density (1).

The energy density corresponds to the cosmological
constant is
\begin{eqnarray}\label{InfBH}
\rho_{2}(\Lambda)=\frac{1}{8\pi}\Lambda
\end{eqnarray}
and the cosmological constant effective mass within a radius r is
\begin{eqnarray}\label{InfBH}
m_{2}(r,\Lambda)=\int_{0}^{r}4\pi r'^{2}\rho_{2}(\Lambda)dr'=\frac{1}{6}\Lambda r^3.
\end{eqnarray}

According to (4), (10) and (12), we arrive at the relation
\begin{eqnarray}\label{InfBH}
m(r,\Lambda)=\int_{0}^{r}4\pi r'^{2}\rho dr'=\int_{0}^{r}4\pi r'^{2}(\rho_{1}+\rho_{2}) dr'=m_{1}(r,\Lambda)+m_{2}(r,\Lambda)=m_{1}(r,\Lambda)+\frac{1}{6}\Lambda r^3.
\end{eqnarray}

With (13), the metric function $f$ can be putted in the form \cite{Elizabeth,mass1}
\begin{eqnarray}\label{InfBH}
g^{rr}=f=1-\frac{2m(r,\Lambda)}{r}=
1-\frac{2(m_{1}(r,\Lambda)+m_{2}(r,\Lambda))}{r}=1-\frac{2m_{1}(r,\Lambda)}{r}-\frac{\Lambda}{3} r^2.
\end{eqnarray}

Since $\rho_{1}\geqslant0$,
$\frac{d m_{1}(r,\Lambda)}{dr}=4\pi r^{2}\rho_{1}\geqslant0$
and $m_{1}(r,\Lambda)$ is an increasing function of r.
We take the assumption that the total scalar field energy is finite,
which means $m_{1}(r,\Lambda)$ has an upper bound.
Since $m_{1}(r,\Lambda)$ is an increasing and upper bounded function,
the limit value $m_{1}(\infty,\Lambda)$ exists.
As r approaching the infinity, the metric asymptotically goes to
\begin{eqnarray}\label{InfBH}
g^{rr}=f\rightarrow 1-\frac{2m_{1}(\infty,\Lambda)}{r}-\frac{\Lambda}{3} r^2.
\end{eqnarray}
The usual Schwarzschild AdS background satisfies
\begin{eqnarray}\label{InfBH}
g^{rr}= 1-\frac{2M}{r}-\frac{\Lambda}{3} r^2,
\end{eqnarray}
where M is the mass of the spacetime \cite{mass1,mass2,mass3}.
Comparing (15) and (16), we find that $m_{1}(\infty,\Lambda)$ is the mass M,
for similar results in asymptotically flat spacetimes see \cite{ub}.

As approaching the infinity, metric functions are characterized by
\begin{eqnarray}\label{AdSBH}
\chi\rightarrow 0,~~~~~~f\rightarrow -\frac{\Lambda}{3} r^2.
\end{eqnarray}

The finite energy density condition implies that
$m_{1}\thicksim O(r^3)$ and $m_{2}\thicksim O(r^3)$, which are valid near the origin.
And near the origin, asymptotically behaviors of functions are \cite{ng1}
\begin{eqnarray}\label{AdSBH}
\chi'\rightarrow 0,~~~~~~f\rightarrow 1+O(r^2).
\end{eqnarray}

The scalar field equation is
\begin{eqnarray}\label{BHg}
\psi''+(\frac{2}{r}-\frac{\chi'}{2}+\frac{f'}{f})\psi'-\frac{\dot{V}}{f}\psi=0.
\end{eqnarray}

Near the origin, the scalar field equation can be expressed as
\begin{eqnarray}\label{BHg}
\psi''+\frac{2}{r}\psi'-\dot{V}\psi=0,
\end{eqnarray}
which has a regular singular point $r=0$.
According to Frobenius theorem \cite{mr5}, one solution behaves as
\begin{eqnarray}
\psi(r\rightarrow 0)\thicksim \frac{A}{r},
\end{eqnarray}
where A is a nonzero constant.
Near the origin $r=0$, the finite mass condition $M=m_{1}(\infty,\Lambda)=\int_{0}^{\infty}4\pi r'^{2}\rho_{1}(r',\Lambda)dr'<\infty$
requires that $r^2\rho_{1}$ increases slower than $\frac{1}{r}$ as $r\rightarrow 0$. It yields the relation
\begin{eqnarray}\label{InfBH}
r^3\rho_{1}(r,\Lambda)\rightarrow 0~~~for~~~r\rightarrow 0.
\end{eqnarray}
For the solution satisfying (21), there is
$\rho_{1}\geqslant f(\psi')^2\sim (\psi')^2\sim \frac{A^2}{r^4}$ as $r\rightarrow 0$,
which is in contradiction with the relation (22).
Here we use the finite mass condition
to rule out the solution (21) while Hod used the finite energy density
condition to rule out this type of solutions \cite{ng1}.
Around $r=0$ , another physical solution of (20) can be expanded as \cite{ng1}
\begin{eqnarray}\label{BHg}
\psi(r)=a[1+\frac{1}{6}\dot{V}(a^2)\cdot r^2]+O(r^3),
\end{eqnarray}
where $a=\psi(0)$ is the value of the scalar field at the origin.

At the infinity, we impose the vanishing condition for the scalar field as
\begin{eqnarray}\label{InfBH}
&&\psi(\infty)=0.
\end{eqnarray}
As is well known, the AdS boundary usually provides the confinement of the system and serves as a box \cite{box1,box2,box3,box4,box5}.
There is freedom in the choice of the field's behavior at the box/AdS boundary,
such as Dirichlet Boundary Conditions and Robin Boundary Conditions \cite{RB1,RB2}.
So a more general infinity boundary condition
compatible with finite energy may exist.
However, the vanishing condition is essential in the present proof.
As such, the no-go theorem in this work applies to a scalar field vanishing at infinity,
but the possible boson star with more generic boundary conditions is not excluded.

In the case of $a=\psi(0)=0$, the scalar field must have one extremum point $r_{peak}$ \cite{Hod-6}.
At this extremum point, there are following characteristic relations
\begin{eqnarray}\label{InfBH}
\{ \psi^2>0,~~~~\psi\psi'=0~~~~and~~~~\psi\psi''\leqslant0\}~~~~for~~~~r=r_{peak}.
\end{eqnarray}
At $r=r_{peak}$, we arrive at the inequality in the form
\begin{eqnarray}\label{BHg}
\psi\psi''+(\frac{2}{r}-\frac{\chi'}{2}+\frac{f'}{f})\psi\psi'-\frac{\dot{V}}{f}\psi^2<0.
\end{eqnarray}

For $a>0$, there are relations $\psi'(0)=0$ and $\psi''(0)=\frac{1}{3}a\dot{V}(a^2)>0$
implying $\psi' > 0$ for $r>0$ around the origin.
Also considering $\psi(\infty)=0$, the scalar field firstly increases to be more positive
and finally approaches zero at the infinity.
So we conclude that one extremum point $\tilde{r}_{peak}$
of the scalar field must exist. There are following characteristic relations
 \begin{eqnarray}\label{InfBH}
\{ \psi>0,~~~~\psi'=0~~~~and~~~~\psi''\leqslant0\}~~~~for~~~~r=\tilde{r}_{peak}.
\end{eqnarray}
At $r=\tilde{r}_{peak}$, the characteristic inequality is
\begin{eqnarray}\label{BHg}
\psi''+(\frac{2}{r}-\frac{\chi'}{2}+\frac{f'}{f})\psi'-\frac{\dot{V}}{f}\psi<0.
\end{eqnarray}

And in cases of $a<0$, one can deduce the conclusion
that one extremum point $\bar{r}_{peak}$ exists. At this extremum point,
there are following characteristic relations
\begin{eqnarray}\label{InfBH}
\{ \psi<0,~~~~\psi'=0~~~~and~~~~\psi''\geqslant0\}~~~~for~~~~r=\bar{r}_{peak}.
\end{eqnarray}
At $r=\bar{r}_{peak}$, there is the characteristic inequality
\begin{eqnarray}\label{BHg}
\psi''+(\frac{2}{r}-\frac{\chi'}{2}+\frac{f'}{f})\psi'-\frac{\dot{V}}{f}\psi>0.
\end{eqnarray}

Relations (26), (28) and (30) are in
contradiction with the scalar field equation (19).
It means that asymptotically AdS
spherically symmetric regular boson stars
cannot be constructed with static scalar fields,
whose potential is positive semidefinite
and monotonically increases with respect to its argument.

\section{Conclusions}

We studied the gravity model of static massive scalar fields
in the background of spherically
symmetric gravity with negative cosmological constants.
We considered self-gravitating scalar fields vanishing at infinity.
The scalar field potential is positive semidefinite
and monotonically increases as a function of its argument.
We obtained the scalar field characteristic
relations (26), (28) and (30) at extremum points.
However, these characteristic relations are in contradiction
with the static scalar field equation (19),
which means the existence of no-go theorem for static boson stars.
In summary, we found that spherically symmetric
regular boson stars cannot be made of
static scalar fields in the asymptotically AdS background.
We pointed out that the no-go theorem in this work applies to a scalar field vanishing at infinity.
We plan to examine whether there is no-go theorem for more generic boundary conditions
in the next work.

\begin{acknowledgments}

We would like to thank the anonymous referee for the constructive suggestions to improve the manuscript.
This work was supported by the Shandong Provincial Natural Science Foundation of China under Grant
No. ZR2018QA008. This work was also supported by a grant from Qufu Normal University
of China under Grant No. xkjjc201906.

\end{acknowledgments}

\end{document}